\documentclass[preprintnumbers,axodraw,aps]{revtex4}

\usepackage{epsfig}
\usepackage{axodraw}
\def\lsim{\raise0.3ex\hbox{$\;<$\kern-0.75em\raise-1.1ex\hbox{$\sim\;$}}}
\def\gsim{\raise0.3ex\hbox{$\;>$\kern-0.75em\raise-1.1ex\hbox{$\sim\;$}}}

\def\unit{\leavevmode\hbox{\small1\kern-3.6pt\normalsize1}}
 \normalsize
\newcommand{\bmat}{\left(\begin{array}}
\newcommand{\emat}{\end{array}\right)}
\newcommand{\be}{\begin{equation}}
\newcommand{\ee}{\end{equation}}
\newcommand{\bea}{\begin{eqnarray}}
\newcommand{\eea}{\end{eqnarray}}

\def\e{\epsilon}

\def\D{\Delta}

\begin{document}
\preprint{SHEP-12-22}
\title{Smooth Hybrid Inflation and Non-Thermal Type II
Leptogenesis}
\author{Shaaban Khalil$^{1,2,3}$, Qaisar Shafi$^4$, and Arunansu Sil$^5$}
\vspace*{0.3cm}
\affiliation{$^1$Centre for Theoretical Physics, Zewail City of Science and Technology, Sheikh Zayed, 12588, Giza, Egypt.\\
$^2$School of Physics and Astronomy, University of Southampton,
Highfield, Southampton SO17 1BJ, UK.\\
$^3$Department of Mathematics, Faculty of Science, Ain
Shams University,  Cairo, 11566, Egypt.\\
$^4$Bartol Research Institute, Department of Physics
and Astronomy, University of Delaware, Newark, DE 19716, USA.\\
$^5$Department of Physics, Indian Institute of
Technology, Guwahati, Assam 781039, India.}
%\date{\today}

\begin{abstract}
\noindent We consider a smooth hybrid inflation scenario based on
a supersymmetric $SU(2)_L \times SU(2)_R \times U(1)_{B-L}$ model.
The Higgs triplets involved in the model play a key
role in inflation as well as in explaining the observed baryon
asymmetry of the universe. We show that the baryon asymmetry
can originate via non-thermal triplet leptogenesis from the
decay of $SU(2)_L$ triplets, whose tiny vacuum expectation values
also provide masses for the light neutrinos.
\end{abstract}
\maketitle
%%%%%%%%%%%%%%%%%%%%%%%%%%%%%%%%%%

%%%%%%%%%%%%%%%%%%%%%%%%%%%%%%%%%%%%%%% Section I %%%%%%%%%%%%%%%%%%%%%%%%%%%%%%%%%%%%%%%%
\section{{\large \bf Introduction}}

There exists an attractive class of supersymmetric models in which
inflation is closely linked to the supersymmetric grand
unification scale \cite{hybrid1, cope, hybrid2, hybrid3}. Among these
models, supersymmetric hybrid inflation (with minimal K\"ahler
potential) predicts a scalar spectral index close to 0.985 \cite{hybrid1},
to be compared with $n_s = 0.968 {\pm 0.014}$ presented by WMAP7 \cite{wmap7}.
Smooth hybrid inflation, a variant of supersymetric hybrid inflation,
yields a spectral index of 0.97 if supergravity effects 
are ignored.
However, inclusion of supergravity corrections with minimal K\"ahler potential 
leads to higher values of the spectral index even in this case \cite{ns}. 
It has been shown
in \cite{gil, Rehman} that the predicted scalar spectral index in smooth hybrid
inflation model is affected if the non-minimal terms in the K\"ahler potential
are switched on, and $n_s$ close to the WMAP prediction is easily realized. 
For supersymmetric hybrid inflation with soft terms, it is also possible to
reduce $n_s$ to 0.968 \cite{redux}.

Inflation in these models is naturally followed by
leptogenesis \cite{lepto}. Type I leptogenesis from the decay of right
handed neutrinos has been discussed in some details in
recent papers \cite{detail}, where the light neutrino masses are
obtained from type I seesaw. Care has to be exercised to
ensure that leptogenesis is consistent with constraints
that may arise from the observed solar and atmospheric neutrino
oscillations \cite{exp}. Light neutrino masses can also arise from
the so-called type II seesaw mechanism \cite{seesaw} in which
heavy scalar $SU(2)_L$ triplets acquire tiny vacuum expectation
values (vevs) that can contribute to the masses of the observed
neutrinos.

An interplay between type I and type II seesaw in the generation of 
light neutrino masses \cite{MS}
is also a possibility (for example, while considering a left-right 
symmetric model). If the right handed neutrinos all have superheavy 
masses comparable to $M_{GUT}$ = $O(10^{16} ~\rm GeV)$ or close to it, 
the type I seesaw contribution to neutrino masses alone would be 
too much small to be compatible with the neutrino oscillation data. 
A situation similar to this is adopted in this paper where the triplet 
vev is the main source of light neutrino masses. It is well known that 
these triplet scalars can play an additional important
role by producing the desired lepton asymmetry \cite{leptrip1, leptrip2}. 
They could be present in the early universe from the decay of the inflaton, 
and their own subsequent decay can lead to leptogenesis.

We implement this scenario (type II leptogenesis with smooth hybrid inflation) within a
supersymmetric version of the well known gauge symmetry $SU(2)_L \times SU(2)_R \times
U(1)_{B-L}$ \cite{LR}. (Generalizations to other (possibly larger) gauge symmetries seems
quite plausible.) We restrict our attention to non-thermal leptogenesis which is quite
natural within an inflationary setting. (For type II thermal leptogenesis see \cite{hambye-goran, sking}). We work
here in the framework of smooth hybrid inflation \cite{smooth1, smooth2, ns} , taking into
account the corrections arising from non-minimal K\"ahler potential. To make the scenario as
technically natural as possible, we impose some additional symmetries including a $U(1)_R$ symmetry
\cite{hybrid1}. We find that the constraints from neutrino oscillations as well as leptogenesis can be satisfied with natural values of the appropriate couplings.

%%%%%%%%%%%%%%%%%%%%%%%%%%%%%%%%%%%%%%% Section II %%%%%%%%%%%%%%%%%%%%%%%%%%%%%%%%%%%%%%%%
\section{Higgs Triplets in Left-Right Model}

The quark and lepton superfields have the following transformation
properties under the gauge group $SU(3)_c \times SU(2)_L \times
SU(2)_R \times U(1)_{B-L}$ \cite{LR}:
$$Q = (3,2,1,{1 \over 3}),~~~ Q^c = (3^*,1,2,-{1 \over 3}),~~~
L = (1,2,1,-1), ~~~ L^c = (1,1,2,1).$$
\noindent The Higgs sector consists of
\begin{eqnarray*}
H &=& (1,2,2,0),~~~~~ \Delta^a_L = (1,3,1,2), ~~~~ \bar \Delta^a_L
=
(1,3,1,-2),~~  a = 1,2,\\
\Delta_R &=& (1,1,3,-2), ~~~ \bar \Delta_R = (1,1,3,2).%
\end{eqnarray*}%
Our primary goal, as stated earlier, is the implementation of
non-thermal type II leptogenesis, and to realize it we consider
two pairs of triplets $\D_L$, $\bar \D_L$ (indicated by index $a = 1,2$)
which, through mixing, can produce the CP violation necessary for
generating an initial lepton asymmetry\footnote{ We do not insist
on a completely left-right symmetric Higgs sector. Thus, only one
pair of $\D_R, \bar \D_R$ is considered.}. The model also
possesses a gauge singlet superfield $S$ which plays a vital role
in inflation.

The superpotential is given by:%
\bea%
W &=& S \Bigl [\frac{(\D_R \bar \D_R)^2}{M^2_{S}} - M_X
^2\Bigr ] + \frac{\alpha_{ab}}{M_{S}} \D^a_L \bar \D^b_L \D_R
\bar \D_R
      + \frac{\gamma^a}{M_{S}} H H \bar \D^a_L \bar \D_R
      + f^a_1 L L \D^a_L +
\nonumber\\
&& f_2 L_c L_c \D_R + Y^l L L_c H + Y^q Q Q_c H ,
\label{w}
\eea
where $a, b = 1,2$, and the $SU(2)$, generation and color indices
are suppressed. $M_X$ is a superheavy mass scale
and $M_S$ is the cutoff scale which controls the non-renormalizable terms in
the superpotential. We take the matrix $\alpha_{ab}$ to be real and
diagonal ($\alpha_{ab} = \delta_{ab} \alpha_a$) in our
calculation for simplicity. The first two terms (in the square bracket) are
responsible for inflation. The importance of the remaining terms will be
discussed later in connection with the inflaton decay, reheating,
leptogenesis and neutrino mass generation. A $Z_2$ symmetry along
with $U(1)_R$ global symmetry is imposed in order to realize the
above superpotential. The charges of all the superfields
are listed in Table I. The inclusion of the $Z_2$
symmetry forbids terms like $\D^a_L \bar \D^b_L$ in the superpotential,
but allows the term $\D^a_L \bar \D^b_L \D_R \bar \D_R$. This
ensures that the $SU(2)_L$ triplets are lighter than the superheavy
right handed neutrinos. Apart from its importance in realizing inflation
(would be discussed in the next section), the global $R$-symmetry
plays another important role in our analysis. Its unbroken $Z_2$
subgroup acts as `matter parity', which implies a stable LSP,
thereby making it a plausible candidate for dark matter.
We see from Table I that baryon number violating superpotential
couplings $QQQ$, $Q_cQ_cQ_c$ and $QQQL$ are forbidden by the
$U(1)_R$ symmetry. This also holds for the higher dimensional
operators, so that the proton is essentially stable \cite{proton}.

\begin{table}[h]
\begin{center}
\begin{tabular}{|c|c|c|c|c|c|c|c|c|c|c|}
\hline Charges & $S$ & $\D^a_L$ & $\bar \D^a_L$ & $\D_R$ & $\bar
\D_R$ & $H$ &
$L$ & $L_c$ & $Q$ & $Q_c$ \\
\hline \hline $R$ & 2 & 2 & 0 & 0 & 0 & 1 & 0 & 1 & 0 & 1 \\
\hline
$Z_2$ & 1 & 1 & -1 & 1 & -1 & 1 & 1 & 1 & 1 & 1 \\
\hline
\end{tabular}
\end{center}
%\begin{description}
\caption{\small $R$ and $Z_2$ charges of
superfields.}
%\end{description}
\end{table}

\section{Smooth Hybrid Inflation}

The superpotential term responsible for inflation is
given by
\be
W_{inf} = S \Bigl [\frac{(\D_R \bar \D_R)^2}{M^2_{S}} - M_X
^2\Bigr ]. \label{1}
\ee
Note that under $U(1)_R$, $S$ carries the same charge as
$W$ and therefore guarantees the linearity of the superpotential
in $S$ to all orders (thus excluding terms like $S^2$ which
could ruin inflation \cite{hybrid1}.). The scalar potential
derived from $W_{inf}$ is %
\be %
V_{inf} = \Bigl |\frac{(\D_R \bar \D_R)^2}{M^2_S} - M_X^2 \Bigr
|^2 + 4|S|^2\frac{ |\D_R|^2 |\bar \D_R|^2 |}{M^4_S} \bigl
(|\D_R|^2 + |\bar \D_R|^2 \bigr ) + D ~\rm{terms}. \label{v} \ee
Using the $D$-flatness condition $|\langle \D_R \rangle | = |
\langle \bar \D_R \rangle |$, we see that the supersymmetric
vacuum lies at $M$ = $|\langle \D_R \rangle| = |\langle \bar \D_R
\rangle| = \sqrt{M_X M_S}$ and $\langle S \rangle = 0$. Defining
$\zeta/2 = |\D_R^0| = |\bar \D_R^0|$ and  $\sigma/{\sqrt{2}} =
|S|$, one can rewrite the scalar potential as \cite{smooth1,
smooth2} %
\be %
V_{inf} = \Bigl [{\frac{\zeta^4}{16 M^2_S}} - M_X^2 \Bigr ]^2 +
\frac{{\sigma}^2 \zeta^6} {16 M^4_S}. \ee The importance of this
potential in the context of inflation is discussed in
\cite{smooth2}. Here we can briefly summarize it. Although $\zeta$
= 0 is a flat direction, it is actually a point of inflection with
respect to any value of $\sigma$. It also possesses two
(symmetric) valleys of local minima (containing the supersymmetric
vacua) which are suitable for inflation. Unlike `regular'
supersymmetric hybrid inflation, the inclination of these valleys
is already non-zero at the classical level and the end of
inflation is smooth.

%%%%%%%%%%%%%%%%%%%%%%%%%%%%%%%%%%%%%%%%%%%%%%%
%\begin{figure}[t]
%\begin{center}
%\hskip -2 cm
%\includegraphics[angle=0,width=10.5cm]{nmsmooth2r.eps}
%\end{center}
%\vskip -0.7 cm \caption{Spectral index $n_s$ versus the symmetry
%breaking scale $M$.}
%\end{center}
%\label{fig1}
%\end{figure}
%%%%%%%%%%%%%%%%%%%%%%%%%%%%%%%%%%%%%%%%%%%%%%%%%%%%%%%

If we set $M = M_{GUT} = 2.86 \times 10^{16}$ GeV, and substitute
in the expression for the quadrupole anisotropy, $(\delta T/T)_Q$,
we find $M_X \simeq 1.8 \times 10^{15}$ GeV and $M_S \simeq 4.6
\times 10^{17}$ GeV \cite{ns}. Here we have employed WMAP7 \cite{wmap7},
measurement of the amplitude of curvature perturbation ($\Delta_{\cal{R}}$) and set the number of
$e$-foldings $N_Q \simeq 57$.
The value of $\sigma$ is $1.3 \times 10^{17}$ GeV at the end of
inflation (corresponding to the slow roll violating parameter,
$\eta = \frac{M^2_P V^{''}}{8\pi V} = - 1$), and it is $2.7
\times 10^{17}$ GeV ~$(\sigma_Q)$ at the horizon exit. The spectral
index is estimated to be $n_s \simeq 0.97$ (without supergravity corrections), 
close to the value of $n_s$ from WMAP7.

Note that the supergravity corrections are important and this
is studied in \cite{ns}. Once these are included (with minimal K\"ahler potential),
$n_s$ approaches unity (for $M \gtrsim 1.5 \times 10^{16}$ GeV)
\cite{ns}. By lowering the scale $M$ compared to the $M_{GUT}$, one can
achieve $n_s$ in the acceptable range. However, in this case the inflaton
field-value $\sigma_Q$ would be larger than the cutoff scale $M_{S}$ providing a threat to the effective field theory concept.

If we employ a non-minimal K\"ahler potential
\be
K = |S|^2 + |\D_R|^2 + |\bar \D_R|^2 + \frac{\kappa_S}{4}
\frac{|S|^4}{M^2_p},
\ee
then along the D-flat direction $|\D_R| =|\bar \D_R|$, the
inflationary potential for $\sigma^2 \gg M^2$ isgiven by,
\be
V = M_X^4 \Bigl [1 - \kappa_S \frac{\sigma^2}{2 M^2_p} +
\bigl (1 - \frac{7}{2} \kappa_S + 2 \kappa^2_S \bigr ) \frac{\sigma^4}{8
M^4_p} - \frac{2}{27}\frac{M^4}{\sigma^4} \Bigr ]. \label{vsugra}
\ee
The spectral index calculated from this potential is in the
desired range ($0.968\pm0.014$) for different choices of $\kappa_S$. 
An analysis of this case is
extensively studied in \cite{Rehman}. We have tabulated sets
of values of $M, M_S, \sigma_Q$ in Table II corresponding to
different choices of $\kappa_S$ with different predictions for the
spectral index (for more examples, see Figs. 7 and 8 of \cite{Rehman}).
With non minimal K\"ahler terms included, there arises the
possibility of having observable tensor to scalar ratio $r$, a canonical
measure of gravity waves produced during inflation \cite{gravity-wave}.

\begin{table}[h]
\begin{center}
\begin{tabular}{|c|c|c|c|c|c|}
\hline
Set & $\kappa_S$ & $n_s$ & $M$ (GeV) & $M_S$(GeV) & $\sigma_Q$(GeV)\\
\hline \hline I & $0$ & 0.99 & 1.2 $\times 10^{16}$ & 1.8
$\times 10^{17}$ & 1.8 $\times 10^{17}$  \\
\hline
II & $0.005$ & 0.968 & 2.2 $\times 10^{16}$ & 5.5 $\times 10^{17}$ & 2.1 $\times 10^{17}$ \\
\hline
III & $0.01$ & 0.968 & 4 $\times 10^{16}$ & 1.5 $\times 10^{18}$ & 3 $\times 10^{17}$ \\
\hline
\end{tabular}
\end{center}
%\begin{description}
\caption{\small For a given value of $\kappa_S$, the predicted values of the
spectral index ($n_s$), the gauge
symmetry breaking scale ($M$), the cutoff scale ($M_S$), and the
inflaton field at the time of horizon exit ($\sigma_Q$) are presented.}
%\end{description}
\end{table}

%%%%%%%%%%%%%%%%%%%%%%%%%%%%%%%%%%%%%%%%%%%%%%%%%%
%\begin{figure}[ht]
%\begin{center}
%\hskip -2 cm
%\includegraphics[angle=0,width=10.5cm]{nmsmooth1r.eps}
%\end{center}q
%\vskip -1. cm \caption{$M_S$ and $\sigma_Q$ as functions of gauge
%symmetry breaking scale, $M$.}
%\end{center}
%\label{fig2}
%\end{figure}
%%%%%%%%%%%%%%%%%%%%%%%%%%%%%%%%%%%%%%%%%%%%%%%%

\section{Reheating}

Let us now discuss inflaton decay and reheating. The inflaton
$field(s)$ smoothly enter an era of damped oscillation about the
supersymmetric vacuum. The oscillating system consists of two
scalar fields $S$ and $\theta = (\delta \theta + \delta
{\bar\theta})/{\sqrt{2}}$ ($\delta \theta = \D^0_R - M$ and
$\delta {\bar\theta} = \bar \D^0_R - M$) with a common mass
$m_{inf} = 2 \sqrt{2} M \frac{M^2}{M^2_S}$, which decay into a pair
of left triplets ($\D^a_L, {\bar\D^a_L}$) and their fermionic
partners (${\tilde{\D}}^a_L, {\tilde{{\bar {\D}}}}^a_L$)
respectively through the Lagrangian [see Eq.(2)]
\begin{equation}
L^s = \sqrt{2} \alpha_a \frac{M}{M_S} m_{inf} S^* \D^a_L {\bar
\D^a_L} + h.c.~,~L^{\theta} =  \sqrt{2} \alpha_a \frac{M}{M_S}
\theta {\tilde{\D}^a_L} {\tilde{{\bar {\D}}}^a_L} +
 h.c. .
\end{equation}

\noindent The decay widths of both $S$ and $\theta$ turn out to be%
\be
\Gamma_{inf} = {\frac{3}{4\pi}} \alpha^2_a \Bigl
({\frac{M}{M_S}}\Bigr )^2
 m_{inf}
~ = {\frac{3}{4\pi}} \Bigl ({\frac{M_a}{M}}\Bigr )^2 m_{inf},
\label{decay}
\ee %
where $M_a$ is  the mass of the $SU(2)_L$ triplet given by $M_a =
\alpha_a {\frac{M^2}{M_S}}$ (generated via the non-renormalizable
superpotential coupling $\frac{\alpha_a}{M_S} \D^a_L \bar \D^a_L
\D_R \bar \D_R$, after $\D_R$, $\bar \D_R$ acquire vevs). For this
decay to be kinematically allowed, $\alpha_a \lesssim \sqrt{2}
\frac{M}{M_S}$.
%~(\sim 2.5 \times 10^{-2}$).
The splitting between $M_1$ and $M_2$ (i.e. between $\alpha_1$ and $\alpha_2$) will be
important in estimating the lepton asymmetry. The decay
of inflaton into right-handed neutrinos is kinematically forbidden
since the latter have superheavy mass acquired from the renormalizable
coupling $f_2 L_c L_c \D_R$, with $f_2$ of order unity.

The reheat temperature from the decay of the inflaton
is $T_{R} \simeq \frac{1}{7} \sqrt{\Gamma M_P},$ where
$\Gamma$ represents the total decay width of the inflaton (here it
is $\Gamma_{inf}$), where $M_P = 2.4 \times 10^{18}$ GeV is
the reduced Planck scale. Using the first set of values for $M,
M_S$ specified in Table II, one finds %
\bea%
T_R & \simeq 0.12 \times  \alpha \Bigl
({\frac{M}{M_S}}\Bigr )^2 \sqrt{M M_P} ~{\rm GeV},%
\eea%
where $\alpha = \sqrt{\alpha^2_1 + \alpha^2_2}$. With the
parameters involved in Table II (set II and III), we find $M/M_S
\sim {\cal{O}}(10^{-2})$. Hence the reheat temperature is 
$T_R \sim {\cal{O}} (10^{10 - 11})$ GeV, where the constraint on
$\alpha_a$ is taken into account ($\alpha \sim {\cal{O}} (10^{-3})$). 
Note that such a reheat temperature does not pose any threat if the graviton is
sufficiently heavy \cite{amsb}. Therefore we conclude from the
above discussion that at the end of inflation, the inflaton system
has decayed away into $SU(2)_L$ triplets. We will show in the next
section that the subsequent decay of these $SU(2)_L$ triplets creates
a lepton asymmetry, which is partially converted into the observed baryon
asymmetry via the electroweak sphaleron effects \cite{sphe}.

%%%%%%%%%%%%%%%%%%%%%%%%%%%%%%%%%%%%%%% Section IV %%%%%%%%%%%%%%%%%%%%%%%%%%%%%%%%%%%%%%%%
\section{Type II Non-thermal leptogenesis and Neutrino masses}

In general both the right-handed neutrinos as well
as the left-handed triplets can yield a lepton
asymmetry in left-right models \cite{hambye-goran}. However, in our case with superheavy ($M \sim 10^{16}$ GeV) right handed
neutrinos, the Leptogenesis
would come mainly from the $SU(2)_L$ triplets.
Note that we have considered two pairs of $SU(2)_L$ superfields, so
that the CP asymmetry would be nonzero.

The experimental value of the
baryon to photon ratio is given by
\cite{wmap7}
\be%
\frac{n_B}{n_{\gamma}} \simeq (6.5 \pm 0.4) \times 10^{-10}.%
\ee%
In this respect, the required lepton asymmetry is estimated to be
\cite{conver}
\be%
\left\vert\frac{n_L}{s} \right\vert \simeq (2.67 - 3.02) \times
10^{-10}.%
\ee%  \
To estimate the lepton asymmetry we follow the analysis of ref
\cite{leptrip2}. The Higgs triplet $\D^a_L$ decays into $LL$ and
$HH$ (see Fig. 1(a)), while $\bar \D^a_L$ decays into $\tilde L
\tilde L$ and $\tilde H \tilde H$. The amount of CP violation in
these decays is controlled by the interference of the tree level
process with one-loop diagram (see Fig. 1(b)) as described in
\cite{leptrip2}.
%The diagrams in the decay of their superpartners can be
%similarly constructed and because of supersymmetry, their
%contribution to the CP violation are same corresponding to
%the scalar triplets.

The effective mass-squared matrix of the scalar triplets $\D^a_L$
and $\bar \D^a_L$ is \cite{leptrip2},
%\be
$\D^{a \dag}_L  (\it{M^2})_{ab} \D ^b_L + {\bar \D^{a \dag}_L}
(\it M^{'2})_{ab} {\bar \D ^b}_L$,
%\ee
where \be {\it{M^2}} = \left(\begin{array}{cc}
M^2_1 - i\Gamma_{11}M_1 & -i\Gamma_{12}M_2\\
 -i\Gamma_{21}M_1 & M^2_2 - i\Gamma_{22}M_2
\end{array}
\right), \label{masssq} \ee and ${\it{M^{'2}}}$ has a similar
pattern with $\Gamma_{ab}$ replaced by $\Gamma'_{ab}$. The
contributions to $\Gamma_{ab}$ ($\Gamma'_{ab}$) come from the absorptive part of
the one loop self-energy diagrams for $\D^a_L \rightarrow \D^b_L$
($\bar \D^a_L \rightarrow \bar \D^b_L$), \bea \Gamma_{ab}M_b &=&
{\frac{1}{8\pi}}[\Sigma_{ij}({f^{a*}_{1ij}}{f^b_{1ij}}) p^2_{\D_L}
+ M_a M_b g^a g^{b*}]~,
\nonumber\\
\Gamma'_{ab}M_b &=&
{\frac{1}{8\pi}}[\Sigma_{ij}({f^a_{1ij}}{f^{b*}_{1ij}})M_a M_b +
p^2_{\bar \D_L} g^{a*} g^b], \eea where $i,j$ are generation
indices, $g^a = \gamma^a (\frac{M}{M_S})$ and $p^2_{\D}$ is the
momentum squared of the incoming or outgoing
particle. The physical states
$\chi^{1,2}_+$, $\xi^{1,2}_+$ (with masses $\sim M_{1,2}$) can be
obtained\footnote{The physical states $\chi^{1,2}_-$
($\xi^{1,2}_-$) are also similarly obtained by diagonalizing the
matrix in Eq.(\ref{masssq}) with $\Gamma_{12}$ replaced by
$\Gamma^*_{12}$ and vice versa.} by diagonalizing $\it M^2$, $\it
M'^2$. Here we neglect terms of order
$[{\frac{\Gamma_{ij}M_j}{M^2_1 - M^2_2}}]^2$.

\begin{figure}[t]
\begin{center}\begin{picture}(300,100)(0,0)
\DashArrowLine(30,50)(70,50){2}
\ArrowLine(100,20)(70,50)
\ArrowLine(100,80)(70,50)
\Text(50,60)[t]{$\small \Delta^a_L$}
\Text(95,65)[l]{$\small L_i (H)$}
\Text(95,35)[l]{$\small L_j (H)$}

\DashArrowLine(150,50)(180,50){2}
\DashArrowArcn(200,50)(20,180,360){3}
\DashArrowArc(200,50)(20,180,360){3}
\DashArrowLine(220,50)(250,50){2}
\ArrowLine(280,80)(250,50)
\ArrowLine(280,20)(250,50)

\Text(160,63)[t]{$\small \Delta^a_L$}
\Text(182,75)[t]{$\small H$}
\Text(182,25)[b]{$\small H$}
\Text(235,63)[t]{$\small \Delta^b_L$}
\Text(280,65)[r]{$\small L_i$}
\Text(280,35)[r]{$\small L_j$}

\Text(90,5)[b]{(a)}
\Text(200,5)[b]{(b)}
\end{picture} \\
\caption{$(a)$ Tree level decay(s) of $\Delta_L$ into leptons
(Higgs). $(b)$ One loop self energy diagram for the generation of CP
asymmetry.}
\end{center}
\end{figure}
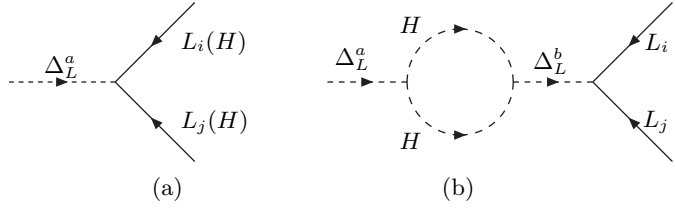

%To obtain the physical states we diagonalize the mass-squared
%matrices. The physical states $\chi^{1,2}_+$, ($\xi^{1,2}_+$)
%with masses $\sim M_1, M_2$
%respectively{\footnote{Here we neglect terms of order
%$[{\frac{\Gamma_{ij}M_j}{M^2_1 - M^2_2}}]^2$ and take $\Gamma_{aa}
%\ll M_a$.}} corresponding to $\it M^2$ ($\it M'^2$) are related to
%$\D^1_L, \D^2_L$ ($\bar \D^1_L, \bar \D^2_L$) by
%\bea
%\left(\begin{array}{c} \D^1_L \\  \D^2_L\end{array}
%\right) = \left(\begin{array}{c} \chi^1_+ + {\frac{i\Gamma_{12}M_2}{M^2_1 - M^2_2}}\chi^2_+\\ \chi^2_+ - {\frac{i\Gamma^*_{12}M_2}{M^2_1 - M^2_2}}\chi^1_+
%\end{array}
%\right).
%\eea

%The physical states $\chi^{1,2}_+$, $\xi^{1,2}_+$ (with masses $\sim M_1, M_2$)
%can be obtained by diagonalizing $\it M^2$, ($\it M'^2$).

%The physical states $\chi^{1,2}_-$ are similarly related to
%$\D^{1*}_L, \D^{2*}_L$, with
%$\Gamma_{12}$ replaced by $\Gamma^*_{12}$ and vice versa.
%A notable fact is that the states $\chi^1_+, \chi^2_+$ are not the
%hermitian conjugates of $\chi^1_-, \chi^2_-$ reflecting
%CP violation{\footnote{The orthonormality conditions are still maintained.}}.

The CP asymmetries are then defined by \cite{leptrip2} \bea \e^a
&=& \triangle L {\frac{\Gamma(\chi^a_- \rightarrow ll) -
\Gamma(\chi^a_+ \rightarrow l^c l^c)}{\Gamma_{\chi^a_-} +
\Gamma_{\chi^a_+}}},
\nonumber\\
 &=& {\frac{ M_1 M_2}{2\pi(M^2_1 - M^2_2)}} {\frac{{\Sigma_{ij}}{\rm{Im}} {f^1_{1ij}}
{f^{2*}_{1ij}} g^1 g^{2*}}{{\Sigma_{ij}} |{f^a_{1ij}}|^2 +
|g^a|^2}}, \eea and \bea \e^{'a} &=& \triangle L
{\frac{\Gamma(\xi^a_+ \rightarrow ll) - \Gamma(\xi^a_- \rightarrow
l^c l^c)}{\Gamma_{\xi^a_+} + \Gamma_{\xi^a_-}}},
\nonumber\\
 &=& {\frac{ M_1 M_2 }{2\pi(M^2_1 - M^2_2)}} {\frac{{\Sigma_{ij}} {\rm{Im}} {f^1_{1ij}}
{f^{2*}_{1ij}} g^1 g^{2*}}{{\Sigma_{ij}} |{f^a_{1ij}}|^2 +
|g^a|^2}}, \eea where the lepton number violation $\D L$ changes by 2
units. We note that $\e^{a}= \e^{'a}$.

The lepton asymmetry is given by
\bea \frac{n_L}{s} & \simeq & \frac{3}{2}
\frac{T_R}{m_{inf}} \Sigma_a 3[\e^{a} + \e^{'a}],
\nonumber \\
& = & \Sigma_a \frac{3}{2} \frac{T_R}{m_{inf}} {\frac{3 M_1 M_2
}{\pi(M^2_1 - M^2_2)}} {\frac{{\Sigma_{ij}} {\rm{Im}} {f^1_{1ij}}
{f^{2*}_{1ij}} g^1 g^{2*}}{{\Sigma_{ij}} |{f^a_{1ij}}|^2 +
|g^a|^2}}, \label{e} \eea where the ratio of the number density of
the $SU(2)_L$ triplets ($n_{\D}$) to the entropy density $s$ is
expressed as $\frac{3}{2} \frac{T_R}{m_{inf}}$. Once this
asymmetry is created, one should ensure that it is not erased by
the lepton-number non-conserving interactions (for example $HH
\leftarrow \D_L \rightarrow LL$, $\tilde H \tilde H \leftarrow
\bar \D_L \rightarrow \tilde L \tilde L$). As long as the
$SU(2)_L$ triplet masses ($M_a$) are sufficiently larger than
$T_R$ (here $\frac{T_R}{M_a} \simeq 0.12 \frac{\sqrt{M
M_P}}{M_S} $ with the specific choice of $M, M_S$ as
given in Table II (set II and III), there will be no significant wash-out
factor, unlike thermal leptogenesis.
%The total lepton asymmetry produced by the decay of these superfields
%is given by
%\be
%{n_L \over {s}} = \frac{3 T_R}{2 m_{inf}} \Sigma_a [\e^{a} +  \e^{'a}].
%\label{e}
%\ee

To estimate $n_L/s$, we need to fix some parameters appearing in
Eq.(\ref{e}) which are also involved in the light neutrino mass
matrix. The neutrino mass matrix is represented by the type II see-saw
relation %
\be%
m_{\nu} = 2{f^a_{1ij}} v^a_{\D_L} - m^T_D M^{-1}_R m_D \equiv
m_{\nu_{II}} - m_{\nu_I},%
\ee%
where $v^a_{\D_L}$ are the $SU(2)_L$ triplet Higgs's vevs. With
the masses of all right handed neutrinos comparable to $M$,
$m_{\nu_I}$ are too small to account for the solar and atmospheric
neutrino data. Hence $m_{\nu_{II}}$ provides the main contribution
to the neutrino mass matrix, namely \be (m_{\nu})_{ij} \simeq 2
{f^a_{1ij}} {\frac{g^a}{M_a}}v^2, \label{mnu} \ee where $v \simeq
174$ GeV. In order to estimate both the lepton asymmetry
(Eq.(\ref{e})) and neutrino masses (through Eq.(\ref{mnu})), we first
simplify by assuming $|g^1| \simeq |g^2| = g, |f^1_1| \simeq
|f^2_1| = f_1 $ (thus $|{\Sigma_{ij}} {f^1_{1ij}} {f^{2*}_{1ij}}|
\simeq \Sigma_{ij} |{f_{1ij}}|^2 $). Then Eqs.(\ref{e}) and
(\ref{mnu}) can be expressed as \bea \frac{n_L}{s} & \simeq &
\frac{9}{\pi} \frac{T_R}{m_{inf}}  \times {\frac{M_1 M_2}{M^2_1 -
M^2_2}}
\times {\frac{\Sigma_{ij} |{f_{1ij}}|^2 g^2}{\Sigma_{ij} |{f_{1ij}}|^2 + g^2}}, \nonumber \\
& \simeq & \frac{0.374}{\pi} \sqrt{\frac{M_P}{M}} \alpha \times
{\frac{M_1 M_2}{M^2_1 - M^2_2}} \times {\frac{\Sigma_{ij}
|{f_{1ij}}|^2 g^2}{\Sigma_{ij} |{f_{1ij}}|^2 + g^2}},
\\
(m_{\nu})_{ij} &\simeq &2 {f_{1ij}} g v^2 \Bigl (\frac{1}{M_1} +
\frac{1}{M_2} \Bigr ), \label{exp} \eea where we have substituted for $T_R$ and $M_a$ and assumed the CP violating phase
to be maximal.

\begin{figure}[t]
\begin{center}
\hskip -2.5 cm \epsfig{file=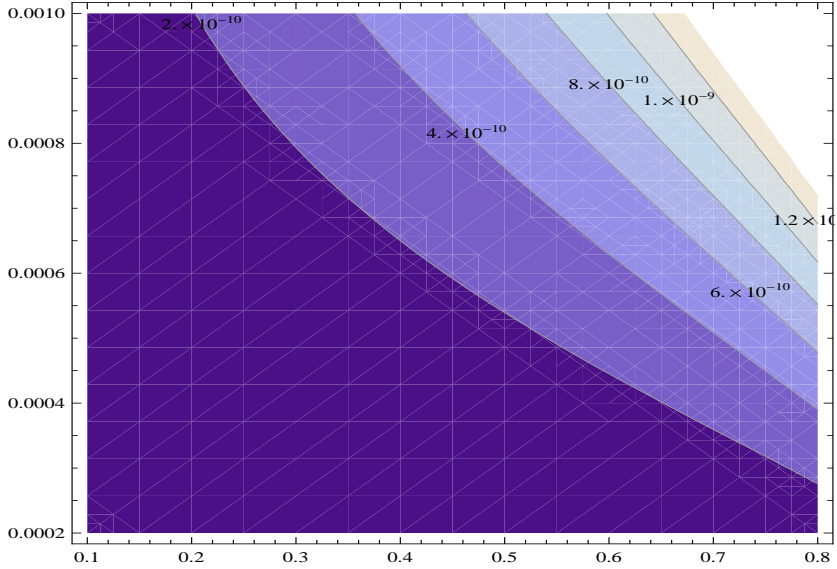,
width=11cm,height=7.5cm,angle=0}
%\includegraphics[angle=0,width=10.5cm]{nL.eps}
%\vskip -6.7 cm
\caption{Contour plot for $n_L/s$ as a function of the parameters:
$p= M_2/M_1$ and $g \lsim M/M_S$.}
\end{center}
\label{nL}
\end{figure}

The neutrino mass matrix $m_{\nu}$ can be diagonalized by \be
m_{\nu} = U^*_{\nu} {m^{diag}_{\nu}} U^{\dag}_{\nu}, \label{mnur}
\ee where ${m^{diag}_{\nu}}$ = diag($m_{\nu_1}, m_{\nu_2},
m_{\nu_3}$).
%$U_{\nu}$ can be
%parametrized by
%\bea
%U_{\nu}&=&\left(
%\begin{array}{ccc} c_{12} c_{13}&s_{12} c_{13}
%& s_{13}
%\\
%-s_{12} c_{23}  -c_{12} s_{23} s_{13}
%  & c_{12} c_{23}  - s_{12}
%s_{23} s_{13}& s_{23} c_{13}
%\\
%s_{12} s_{23}  - c_{12} c_{23}    s_{13}& - c_{12} s_{23} -
%s_{12} c_{23}       s_{13} &   c_{23} c_{13}
%\end{array}
%\right),
%\label{eq:tmix}
%\eea
%where $s_{ij} = \sin\theta_{ij}, c_{ij} = \cos\theta_{ij}$, and we have
%ignored any phases in $U_{\nu}$.
In the basis where the charged lepton matrix is diagonal,
$U_{\nu}$ coincides with the lepton mixing matrix. Using
Eqs.(\ref{exp}),
%and (\ref{mnur}),
we get \bea \frac{n_L}{s} & \simeq &  \frac{0.374}{\pi}
\frac{p\sqrt{1+p^2}}{1-p^2} \sqrt{\frac{M_P}{M}}
\frac{M_1}{M}\frac{M_S}{M} \times \frac{\Sigma_{ij}
|m_{\nu_{ij}}|^2 F g^2}{\Sigma_{ij} |m_{\nu_{ij}}|^2 F + g^4 },
\eea where $F = \frac{M^2_1 M^2_2}{4 v^4 (M_1 + M_2)^2} =
\frac{p^2}{(1+p)^2} \times \frac{M^2_1}{4v^4} $. Here $p$
determines the degree of degeneracy between $M_1$ and $M_2$,
defined by $M_2 = p M_1$. Since the parameter $g$ is defined as $g^a =
\gamma^a \frac{M}{M_S}$, its maximum value is of order
$\frac{M}{M_S}$. Finally, using the current
experimental limits for neutrino masses \cite{fogli-lisi}, one finds that
$\Sigma_{ij} |m_{\nu_{ij}}|^2$
is given by $ \Sigma_{ij}|m_{\nu_{ij}}|^2  \simeq 0.0025 ~ {\rm ~eV}^2,$
where we have used the best fitted values of the neutrino mixing
angles $\theta_{12}, \theta_{23}, \theta_{13}$ and mass squared
differences \cite{fogli-lisi}. We have taken the lightest
neutrino mass eigenvalue to be zero. In Fig. 2 we present the lepton asymmetry
as a function of $p$ and $g$ with $\alpha_1 = 10^{-3}$. We see that
$n_L/s$ can be of order the desired value $(2-3) \times 10^{-10}$
for $ 0.2 \lesssim p \lesssim 0.8$ and $g \gtrsim 2.5 \times 10^{-4}$, which
means $\gamma^a \simeq {\cal O}(0.01)$. It is worth mentioning
that with these values one finds that $M_1$ and $M_2$ are given by
$M_1 \simeq 10^{12}$ GeV and $M_2 \simeq (2-8)\times
10^{11}$ GeV. Therefore, $M_{1,2}/T_R
> 10 $, which indicates that no washout should happen.

%%%%%%%%%%%%%%%%%%%%%%%%%%%%%%%%%%%%%%% Section IV %%%%%%%%%%%%%%%%%%%%%%%%%%%%%%%%%%%%%%%%
\section{Conclusions}

We have considered type II non-thermal leptogenesis in the context
of smooth hybrid inflation. The scheme is consistent with the
observed solar and atmospheric neutrino oscillations. Although our
discussion is based on the gauge symmetry $SU(2)_L \times SU(2)_R
\times U(1)_{B-L}$, it is clear that it could be extended to other
models which contain suitable $SU(2)_L$ triplet scalars with tiny
vevs responsible for the observed neutrino masses.
%The scheme is consistent with the observed solar and
%atmospheric neutrino oscillations
%(Arunansu, this last sentence could be the 2nd sentence???)
The stability of the proton will depend on the underlying gauge
symmetry.\\

\noindent{\bf {Acknowledgements:}}\\

This work was supported in part by U.S DOE under contract number
DE-FG02-12ER41808. The work of S.K. is partially supported by the
Leverhulme Trust under the grant VP2-2011-012. A.S. was supported
by a Marie Curie Fellowship of the European Union Program
MRTN-CT-2004-503369 while a part of this project was carried out.
A.S. also acknowledges the partial support from the Start-Up grant
from IIT, Guwahati.


\begin{thebibliography}{99}

\bibitem{hybrid1} G. Dvali, Q. Shafi and R.K. Schaefer, Phys. Rev. Lett. 73 (1994) 1886.

\bibitem{cope} 
  E.~J.~Copeland, A.~R.~Liddle, D.~H.~Lyth, E.~D.~Stewart and D.~Wands,
  %``False vacuum inflation with Einstein gravity,''
  Phys.\ Rev.\ D  49, 6410 (1994).

\bibitem{hybrid2} G. Lazarides, R.K. Schaefer and Q. Shafi, Phys. Rev. D56 (1997) 1324.

\bibitem{hybrid3} G. Dvali, G. Lazarides and Q. Shafi, Phys. Lett. B424 (1998) 259;
For a review and additional references, see G. Lazarides, Lec. Notes Phys. 592 (2002) 351.

\bibitem{wmap7}
  E.~Komatsu {\it et al.}  [WMAP Collaboration],
  %``Seven-Year Wilkinson Microwave Anisotropy Probe (WMAP) Observations: Cosmological Interpretation,''
  Astrophys.\ J.\ Suppl.\  192, 18 (2011)
  [arXiv:1001.4538 [astro-ph.CO]].

\bibitem{ns} V.~N.~Senoguz and Q.~Shafi, Phys. Lett. B567 (2003) 79.

\bibitem{gil} M. Bastero-Gil, S.F. King and Q.Shafi, hep-ph/0604198.

\bibitem{Rehman}M.~u.~Rehman, V.~N.~Senoguz and Q.~Shafi, Phys. Rev. D75 (2007) 043522.

\bibitem{redux}
  M.~U.~Rehman, Q.~Shafi and J.~R.~Wickman,
  %``Supersymmetric Hybrid Inflation Redux,''
  Phys.\ Lett.\ B 683, 191 (2010).
  %%CITATION = ARXIV:0908.3896;%%

\bibitem{lepto} M. Fukugita and T. Yanagida, Phys. Lett. B174 (1986) 45;
For non-thermal leptogenesis see G. Lazarides and Q. Shafi, Phys.
Lett. B258, (1991) 305.

\bibitem{detail}
G. Lazarides and Q. Shafi, Phys. Rev. D58 (1998) 071702; G.
Lazarides and N.D. Vlachos, Phys. Lett. B441 (1998) 46; R.
Jeannerot, S. Khalil and G. Lazarides, Phys. Lett. B506 (2001)
344; B.~Kyae and Q.~Shafi, Phys. Lett. B556 (2003) 97; Q.~Shafi
and V.~N.~Senoguz, Eur. Phys. J. C33 (2004) S758; V.~N.~Senoguz
and Q.~Shafi, Phys. Lett. B582 (2004) 6; V.~N.~Senoguz and
Q.~Shafi, Phys. Lett. B596 (2004) 8; S.~Dar, Q.~Shafi and A.~Sil,
  Phys.\ Lett.\ B 632, 517 (2006); 
C.~Pallis and N.~Toumbas, arXiv:1207.3730 [hep-ph].

%\bibitem{inflation} A.H. Guth, Phys. Rev. D23 (1981) 347; For reviews,
%A.D.Linde, Rept. Prog. Phys. 47 (1984) 925; G. Lazarides, Lect.
%Notes Phys. 592 (2002) 351; K.A. Olive, Phys. Rept. 190 (1990)
%307.

\bibitem{exp} Super-Kamiokande Collaboration, T. Nakaya, eConf C020620 (2002) SAAT01;
Super-Kamiokande Collaboration, S. Fukuda et. al., Phy. Rev. Let.
89 (2002) 179; SNO Collaboration, Q.R. Ahmad et. al., Phys. Rev.
Let. 89 (2002) 011302; Kamland Collaboration, K. Eguchi et. al.,
Phys. Rev. Let. 90 (2003) 021802; CHOOZ Collaboration, M.
Apollonio et. al., Phys. Let. B466 (1999) 415; G.~L.~Fogli,
E.~Lisi, A.~Marrone, A.~Palazzo and A.~M.~Rotunno,
    %``Neutrino mass and mixing parameters: A short review,''
      [arXiv:hep-ph/0506307].
  %%CITATION = HEP-EX 0212021;%%


\bibitem{seesaw}  G. Lazarides, Q. Shafi and
C. Wetterich, Nucl. Phys. B181 (1981) 287; M. Magg and C.
Wetterich, Phys. Lett. B94 (1980) 61; J. Schechter and J.W.F.
Valle, Phys. Rev. D22 (1980) 2227; R.N. Mohapatra and G.
Senjanovic, Phys. Rev. D23 (1981) 165.

\bibitem{MS}
  E.~K.~Akhmedov and M.~Frigerio,
  %``Interplay of type I and type II seesaw contributions to neutrino mass,''
  JHEP 0701, 043 (2007).

\bibitem{leptrip1} G. Lazarides and Q. Shafi, Phys. Rev. D58 (1998) 071702;
E. Ma and U. Sarkar, Phys. Rev. Lett. 80 (1998) 1171.

\bibitem{leptrip2} T. Hambye, E. Ma and U. Sakar, Nucl. Phys. B602 (2001) 23.

\bibitem{LR} J.~C.~Pati and A.~Salam,
  %``Lepton Number as the Fourth Color,''
  Phys.\ Rev.\ D 10, 275 (1974); R.~N.~Mohapatra and J.~C.~Pati,
  %``Left-Right Gauge Symmetry and an Isoconjugate Model of CP Violation,''
  Phys.\ Rev.\ D 11, 566 (1975); G.~Senjanovic and R.~N.~Mohapatra,
  %``Exact Left-Right Symmetry and Spontaneous Violation of Parity,''
  Phys.\ Rev.\ D 12, 1502 (1975); G.~Senjanovic,
  %``Spontaneous Breakdown of Parity in a Class of Gauge Theories,''
  Nucl.\ Phys.\ B 153, 334 (1979);  
M.~Magg, Q.~Shafi and C.~Wetterich,
  %``Gauge Hierarchy In Presence Of Discrete Symmetry,''
  Phys.\ Lett.\ B 87, 227 (1979); M.~Cvetic,
  %``Spontaneous Breaking Of The Left-right Symmetry And Quantum Corrections,''
  Nucl.\ Phys.\ B 233, 387 (1984);  M. Cvetic and J. Pati, 
Phys. Lett. B135 (1984) 57; R.N. Mohapatra
and A. Rasin, Phys. Rev. D54 (1996) 5835; R. Kuchimanchi, Phys.
Rev. Lett. 76 (1996) 3486; R.N. Mohapatra, A. Rasin and G.
Senjanovic, Phys. Rev. Lett. 79 (1997) 4744; C. Aulakh, K. Benakli
and G. Senjanovic, Phys. Rev. Lett. 79 (1997) 2188; C.S. Aulakh,
A. Melfo and G. Senjanovic, Phys. Rev. D57 (1998) 4174.

\bibitem{hambye-goran}
T. Hambye and G. Senjanovic, Phys. Lett. B582 (2004) 73. 

\bibitem{sking} S.~Antusch and S.~F.~King, Phys. Lett. B597 (2004) 199;
W.~Rodejohann, Phys. Rev. D70 (2004) 073010; W.~l.~Guo, Phys. Rev.
D70 (2004) 053009; N.~Sahu and S.~Uma Sankar, Phys. Rev. D71
(2005) 013006; S.~Antusch and S.~F.~King, JHEP0601 (2006) 117;
T.~Hambye, M.~Raidal and A.~Strumia, Phys. Lett. B632 (2006) 667;
P.~Hosteins, S.~Lavignac and C.~A.~Savoy, Nucl.\ Phys.\  B 
755, 137 (2006);
T.~Hallgren, T.~Konstandin and T.~Ohlsson, %``Triplet Leptogenesis in Left-Right Symmetric Seesaw Models,''
JCAP 0801, 014 (2008).


\bibitem{smooth1} G. Lazarides and C. Panagiotakopoulos, Phys. Rev. D52 (1995)
R559.
\bibitem{smooth2} G.~Lazarides, C.~Panagiotakopoulos and N.~D.~Vlachos,
Phys. Rev. D54, (1996) 1369;  R.~Jeannerot, S.~Khalil and
G.~Lazarides, Phys. Lett. B506, (2001) 344; For an updated
analysis see ref. \cite{ns}.

\bibitem{proton}
  M.~U.~Rehman, Q.~Shafi and J.~R.~Wickman,
  %``Minimal Supersymmetric Hybrid Inflation, Flipped SU(5) and Proton Decay,''
  Phys.\ Lett.\ B 688, 75 (2010).
  %%CITATION = ARXIV:0912.4737;%%

\bibitem{gravity-wave}
  Q.~Shafi and J.~R.~Wickman,
  %``Observable Gravity Waves From Supersymmetric Hybrid Inflation,''
  Phys.\ Lett.\ B 696, 438 (2011);
  %%CITATION = ARXIV:1009.5340;%%
  M.~U.~Rehman, Q.~Shafi and J.~R.~Wickman,
  %``Observable Gravity Waves from Supersymmetric Hybrid Inflation II,''
  Phys.\ Rev.\ D 83, 067304 (2011);
  M.~U.~Rehman and Q.~Shafi,
  %``Simplified Smooth Inflation with Observable Gravity Waves,''
  arXiv:1202.0011.
  %%CITATION = ARXIV:1202.0011;%%

\bibitem{amsb}
  T.~Gherghetta, G.~F.~Giudice and J.~D.~Wells,
  %``Phenomenological consequences of supersymmetry with anomaly induced masses,''
  Nucl.\ Phys.\ B {\bf 559}, 27 (1999)
  [hep-ph/9904378]; T.~Higaki, K.~Kamada and F.~Takahashi,
  %``Higgs, Moduli Problem, Baryogenesis and Large Volume Compactifications,''
  arXiv:1207.2771 [hep-ph].

\bibitem{sphe} V.A. Kuzmin, V.A. Rubakov and M. Shaposhnikov, Phys. Lett. B155
(1985) 36; P. Arnold and L. McLerran, Phys. Rev. D36 (1987) 581.

%\bibitem{conver} S.Y. Khlebnikov, M.E. Shaposhnikov, Nucl. Phys. B308 (1988)
885; J.A. Harvey, M.S. Turner, Phys. Rev. D42 (1990) 3344.

%\bibitem{cos} S.~Hannestad, hep-ph/0409108.

\bibitem{fogli-lisi}
  G.~L.~Fogli, E.~Lisi, A.~Marrone, D.~Montanino, A.~Palazzo and A.~M.~Rotunno,
  %``Global analysis of neutrino masses, mixings and phases: entering the era of leptonic CP violation searches,''
  arXiv:1205.5254 [hep-ph].
\end{thebibliography}
\end{document}